\documentclass[prl,twocolumn,amsmath,amssymb,superscriptaddress,showpacs]{revtex4}
\usepackage{graphics}
\usepackage{epsfig}
\usepackage{bbm}
\usepackage[latin1]{inputenc} 

\newcommand{\be}{\begin{equation}}
\newcommand{\ee}{\end{equation}}
\newcommand{\eea}{\end{eqnarray}}
\newcommand{\bea}{\begin{eqnarray}}

\newcommand{\mean}[1]{\ensuremath{\langle{#1}\rangle}}

\newcommand{\eins}{\ensuremath{\mathbbm 1}}
\newcommand{\qed}{\ensuremath{\hfill \Box}}
\newcommand{\fa}{\ensuremath{\mathfrak{a}}}
\newcommand{\fb}{\ensuremath{\mathfrak{b}}}
\newcommand{\fs}{\ensuremath{\mathfrak{s}}}
\newcommand{\WW}{\ensuremath{\mathcal{W}}}

\newcommand{\CC}{\ensuremath{\mathcal{C}}}
\newcommand{\DD}{\ensuremath{\mathcal{D}}}
\newcommand{\BB}{\ensuremath{\mathcal{B}}}

\newcommand{\ketbra}[1]{\ensuremath{| #1 \rangle \langle #1 |}}

\newcommand{\ket}[1]{\ensuremath{|#1\rangle}}

\newcommand{\braket}[2]{\ensuremath{\langle #1|#2\rangle}}

\newcommand{\kommentar}[1]{}

\begin{document}
\title{Bell Inequalities for Graph States }
\date{\today}
\begin{abstract}
We investigate the non-local properties of graph states. 
To this aim, we derive a family of Bell inequalities 
which require three measurement settings for each party 
and are maximally violated by graph states. 
In turn, for each graph state there is an inequality maximally 
violated only by that state.
We show that for certain types of graph states the violation 
of these inequalities increases exponentially with the number 
of qubits. We also discuss connections 
to other entanglement properties such as the positivity of the 
partial transpose or the geometric measure of entanglement. 
\end{abstract}

\author{Otfried G\"uhne}

\affiliation{Institut f\"ur Quantenoptik und Quanteninformation,
\"Osterreichische Akademie der Wissenschaften,
A-6020 Innsbruck, Austria,}

\author{G\'eza T\'oth}

\affiliation{Max-Planck-Institut f\"ur
Quantenoptik, Hans-Kopfermann-Stra{\ss}e 1, D-85748 Garching, Germany,}

\author{Philipp Hyllus}

\affiliation{Institut f\"ur Theoretische Physik, Universit\"at
Hannover, Appelstra{\ss}e 2, D-30167 Hannover, Germany,}

\affiliation{Institute for Mathematical Sciences, 
Imperial College London, 48 Prince's Gardens, SW7 London 2AZ, UK,}

\author{Hans J. Briegel}

\affiliation{Institut f\"ur Quantenoptik und Quanteninformation,
\"Osterreichische Akademie der Wissenschaften,
A-6020 Innsbruck, Austria,}

\affiliation{Institut f\"ur Theoretische Physik,
Universit\"at Innsbruck, Technikerstra{\ss}e 25,
A-6020 Innsbruck, Austria}



\maketitle

Quantum theory predicts correlations which are 
stronger than the correlations of local hidden 
variable (LHV) models. By definition, LHV models 
have to obey the constraints of realism and 
locality: Any observable has a predetermined 
value, regardless of whether it is measured or 
not, and the choice of which observable to measure 
on one party of a multipartite system does not 
affect 
the results  of
the other parties. These constraints lead 
to the so-called Bell inequalities which put bounds 
on the correlations. These inequalities turn out to 
be violated by certain quantum mechanical states 
\cite{bell64, pesca}. 

In this letter we address the question, whether
graph states allow a LHV description or not. 
Graph states form a family of multi-qubit states 
which comprises many popular states such as the 
Greenberger-Horne-Zeilinger (GHZ) states and the 
cluster states \cite{graphs1}. Graph states 
are also crucial for applications: All 
codewords in the standard quantum error correcting 
codes correspond to graph states \cite{graphapp2}
and one-way quantum computation uses graph 
states as resources \cite{graphapp1}. Recently,
graph states have been produced in optical lattices 
\cite{bloch} and the basic elements of one-way quantum 
computing have been demonstrated experimentally \cite{anton}. 
Also, general methods for the generation of graph states 
have been explored \cite{zang}.

It is a natural and important question whether these 
tasks and experiments, including the effects of noise
can be described by LHV models. To answer this question, 
we derive a class of Bell inequalities. 
Each graph state violates one of these inequalities in the GHZ 
sense, i.e., by saturating all correlation terms and for certain 
types of graph states the violation of local realism increases 
exponentially with the number of qubits. In this way we show that 
tasks like measurement based quantum computing and quantum error 
correction are far from the realm of LHV theories.
Note that the non-locality of special examples of 
graph states has been shown recently \cite{pesca}.

Graph states are defined as follows. Let $G$ be a 
{\it graph,} i.e., a set of $n$ vertices and some 
edges connecting them. Some interesting graphs are 
shown in Fig.~1. For each vertex $i$ the neighborhood $N(i)$ 
denotes the vertices which are connected with $i.$
We can associate to each vertex $i$ a {\it stabilizing operator} 
$g_i$ by
\begin{equation}
g_i=
X^{(i)}\bigotimes\nolimits_{j\in N(i)} Z^{(j)}.
\end{equation}
From now on, 
$X^{(i)}, Y^{(i)}, Z^{(i)}$ denote the 
Pauli matrices $\sigma_x,\sigma_y,\sigma_z,$ 
acting on the $i$-th qubit. For instance, for 
the fully connected three vertex graph, 
the stabilizing operators are
$g_1=X^{(1)}Z^{(2)}Z^{(3)}, g_2=Z^{(1)}X^{(2)}Z^{(3)}$
and $g_3=Z^{(1)}Z^{(2)}X^{(3)}.$ The 
{\it graph state} $\ket{G}$ associated with 
the graph $G$ is the unique $n$-qubit state 
fulfilling
\be
g_i \ket{G}= \ket{G}, \mbox{ for } i=1,...,n.
\ee
Physically, the graph describes the perfect correlations 
in the state $\ket{G},$ since 
$\mean{g_i}=\mean{X^{(i)}\bigotimes_{j\in N(i)} Z^{(j)}}=1.$
At the same time, it denotes a possible interaction history
leading to $\ket{G},$ i.e., $\ket{G}$ can be produced by 
an Ising type interaction acting between the connected qubits.

\begin{figure}[t]

\centerline{\epsfxsize=0.98\columnwidth
\epsffile{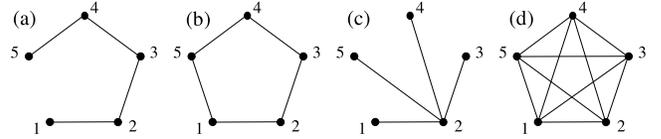} }

\caption{ 
Types of graphs for the case of five vertices: 
(a) The linear cluster graph $LC_5.$
(b) The ring cluster graph $RC_5.$
(c) The star (or GHZ) graph $ST_5.$ 
This describes a GHZ state.
(d) The fully connected graph $FC_5.$
This graph can be obtained from $ST_5$
by local complementation on the second qubit 
(see Lemma 4). It also describes a GHZ state 
\cite{graphs1}.
}
\label{fig_settings}
\end{figure}

Given the  stabilizing operators $g_i,$
we can look at the group of their products, 
the so-called {\it stabilizer} \cite{G96},
\be
S(G)=\{s_j, j=1,...,2^n\}; \;\;\; s_j=\prod_{i \in I_j(G)} g_i
\label{stabdef}
\ee
where $I_j(G)$ denotes a subset of the vertices of $G.$
If a certain generator $g_k$ appears in the product in 
Eq.~(\ref{stabdef}), (i.e., $k\in I_j(G)$), we say that 
{\it $s_j$ contains $g_k.$} The group $S(G)$ is 
commutative and has $2^n$ elements. Of course, 
for all these elements $s_j\ket{G}=\ket{G}$ holds. 
A last property we have to mention is the fact that 
\be
\sum_{i=1}^{2^n} s_i = 2^n \ketbra{G}
\label{stab}
\ee
holds, as can be checked by direct calculation \cite{graphs1}.

Now we present our idea for the derivation of Bell 
inequalities. Given a graph $G$ all the stabilizing 
operators are of the form
\be
s_i(G)=\bigotimes_{k=1}^{n} O^{(k)}_i,
\ee
where the single qubit observables are either the identity or one 
of the Pauli matrices: 
$O^{(k)}_i\in \{\eins, \pm X^{(k)}, \pm Y^{(k)}, \pm Z^{(k)}\}.$ 
We look at the operator
\be
\BB(G)=\sum_i^{2^n} s_i(G)= \sum_i^{2^n}\bigotimes_{k=1}^{n}O^{(k)}_i.
\ee
To give a simple example, this operator for the fully 
connected graph state for three qubits reads
\begin{eqnarray}
\BB(FC_3) &= &\eins^{(1)}\eins^{(2)}\eins^{(3)}
             +X^{(1)}Z^{(2)}Z^{(3)} 
             +Z^{(1)}X^{(2)}Z^{(3)}
\nonumber \\ &             
+&Z^{(1)}Z^{(2)}X^{(3)}
             +Y^{(1)}Y^{(2)}\eins^{(3)}
             +Y^{(1)}\eins^{(2)}Y^{(3)} 
\nonumber \\ &+& 
             \eins^{(1)}Y^{(2)}Y^{(3)}
             -X^{(1)}X^{(2)}X^{(3)}.
\label{example} 
\end{eqnarray}
We take this $\BB$ as the Bell operator and compute a bound
\be
\CC(G) \equiv \CC(\BB) = \max_{\rm LHV}|\mean{\BB}|,
\ee
where the maximum of the absolute value of the 
mean value \mean{\BB} is taken over all LHV models.
Here, it suffices to look at deterministic LHV models 
which have to assign definite values $\{+1,-1\}$ to the 
observables $O^{(i)}_k,$ whenever $O^{(i)}_k \neq \eins$ 
\cite{remark2}. This is due to the fact that nondeterministic 
LHV models can be viewed as deterministic LHV models where 
the hidden variables are not known.
In principle, in the definition of $\BB$ the Pauli 
matrices can be replaced by {\it arbitrary} dichotomic 
observables. Since we are interested in graph states, 
we will, however, always use $X,Y,Z.$

If we can find for a given graph $G$ a bound 
$\CC(G)<2^n,$ the nonlocality of the graph state 
$\ket{G}$ is 
detected. This is due to Eq.~(\ref{stab}), 
which implies that for the graph state 
$\mean{\BB}=2^n$ holds. Also, the graph state  
violates the Bell inequality maximally. 
In the example of Eq.~(\ref{example}) we will see later that
$\CC(FC_3)=6.$ This gives rise to the Bell inequality
$|\mean{\BB(FC_3)}| \leq 6$ which is violated by the state
$\ket{FC_3}.$ 
In the following, it will also be useful to compare 
the strength of the Bell inequalities by the normalized 
parameter $\DD(G):=\CC(G)/2^n$ or $1/\DD.$ Note that
we have a valid Bell inequality whenever $\DD < 1.$

So the main task is to find the value of  
$\CC(G)$ or $\DD(G).$ An exact calculation is, in general, 
very demanding. However, as we will show, it 
is quite easy to obtain bounds on $\DD(G)$ depending 
on the structure of the graph $G,$ especially when 
we can identify some subgraphs where the bounds 
are already known. Together with the exact 
calculation of $\DD(G)$ for graphs with a 
small number of qubits this allows us to derive 
some general results for arbitrary graphs. 
Let us first note two useful facts about the 
dependence of $\DD$ on the LHV models.

{\bf Lemma 1.}
We can restrict our attention to LHV models 
which assign $+1$ to all $Z$ measurements.
\\
{\it Proof.} 
In an element $s_j$ of the stabilizer we have
$O^{(i)}\in \{ Y^{(i)}, Z^{(i)} \}$ iff the number 
of $Y$ and $X$ in $N(i)$ is odd. So if a LHV model
assigns $-1$ to $Z^{(i)}$, we can, by changing 
the signs for $Z^{(i)},Y^{(i)}$ and for all $X^{(k)}$
and all $Y^{(k)}$ with $k\in N(i),$ obtain a 
LHV model with the same mean value of $\BB$ and 
the desired property.
$\qed$

{\bf Lemma 2.} Let $\BB$ be a Bell operator for
an arbitrary graph and let $\BB'$ be  the Bell 
operator which is obtained from $\BB$ by making 
a permutation
\be
P: \{ \eins^{(i)}, X^{(i)}, Y^{(i)}, Z^{(i)}\} 
\rightarrow \{ \eins^{(i)}, X^{(i)}, Y^{(i)}, Z^{(i)}\}
\ee
of the observables on one qubit.  Then $\DD(\BB)=\DD(\BB').$
\\
{\it Proof.} 
It suffices to show the Lemma for a transposition
$O^{(i)}\leftrightarrow \tilde O^{(i)}.$  
Some transpositions 
are a renaming of variables, and the interesting 
transpositions are of the type 
$A^{(i)}\leftrightarrow \eins^{(i)}$ with 
$A^{(i)}=X^{(i)},Y^{(i)},Z^{(i)},$ say $A^{(i)}=X^{(i)}$ 
for definiteness. If a given LHV model $LHV_1$ 
assigns $+1$ to $X^{(i)}$ the transposition 
$X^{(i)}\leftrightarrow \eins^{(i)}$ does not 
change $\mean{\BB}.$ If the LHV model assigns $-1$ 
to $X^{(i)}$ we can construct a new LHV model $LHV_2$ 
from $LHV_1$ by flipping the signs from $Y^{(i)},Z^{(i)}.$ 
This fulfills $\mean{\BB'}_{LHV_1}=-\mean{\BB}_{LHV_2}.$
This proves the claim, since $\CC$ is defined 
via the absolute value.
$\qed$

Now we derive an estimate for $\DD(G),$ when 
$G$ is built out of two other graphs $G_1$ and $G_2$ 
in a certain way.

{\bf Lemma 3.} Let $G_1, G_2$ be two graphs and let 
$G$ be the graph which comprises $G_1$ and $G_2$  
and one single connection between one vertex of $G_1$ 
and one of $G_2,$ i.e.,
$G=G_1 \mbox{ --- } G_2.$
Then
\be
\DD(G)\leq \DD(G_1) \DD(G_2).
\ee
{\it Proof.} The proof is given in the Appendix.
$\qed$

It is much more demanding to derive bounds on 
$\DD(G)$ when $G$ is made out of subgraphs in 
a more complicated way than the way above. 
However, it is easy to see that $\DD(G) < 1$ whenever 
$G$ contains a subgraph $G_1$ with $\DD(G_1)< 1.$
This is due to the fact that the stabilizer of $G_1$
is a subset of the stabilizer of $G$ up to some extra 
$Z$ terms which can be neglected due to Lemma 1. 

Finally, we want to show the invariance of $\DD$ under the 
so-called {\it local complementation} of a graph. This 
transformation acts as follows: One picks up a 
vertex $i_0$ and inverts the neighborhood $N(i_0),$ 
i.e., all connections between two vertices belonging 
to $N(i_0)$ are cut and vertices in $N(i_0)$ which 
were unconnected become connected. Connections between 
$N(i_0)$ and the rest of the graph are not affected. 
To give an example the graph $ST_n$ can be transformed by 
a local complementation on the central qubit into the graph
$FC_n$ (see Fig.~1).
The local complementation of a graph acts on the 
graph state as a local unitary transformation of the 
(local) {\it Clifford group} \cite{graphs1,graphs2}. 
This means that it 
transforms on each qubit Pauli matrices into Pauli 
matrices. So we have: 

{\bf Lemma 4.} Let $G_1$ be a graph and $G_2$ 
be a graph which arises from $G_1$ by local 
complementation. Then $\DD(G_1)=\DD(G_2).$
\\
{\it Proof.} 
Since the local complementation maps Pauli matrices to Pauli 
matrices on each qubit, $\DD$ is not changed
due to Lemma 2. 
$\qed$

After showing in the previous Lemmata how to estimate
$\DD(G)$ when the values of subgraphs are given, it 
is now time to calculate $\DD(G)$ for graphs with a small 
number of qubits $n.$ This can easily be done by a computer, 
by checking $\mean{\BB}$ for all the $8^n$ LHV models. 
In these calculations, Lemma 1 reduces the computational
complexity significantly. The numerical results for 
interesting graphs up to ten qubits are shown in 
Table I.

\begin{table}
\caption{
Value for $\DD(G)$ for different interesting 
graphs (see Fig.~1) and different numbers  
of qubits $n$.
}

\begin{tabular}{|c||c|c|c|c|c|c|c|c|}
\hline
$n$  & 3 & 4 & 5 & 6 & 7 & 8 & 9 & 10
\\
\hline
\hline
${LC_n}$& 3/4 &3/4& 5/8&9/16&8/16&7/16&25/64&22/64
\\
\hline
${RC_n}$& {3}/{4} & 3/4&5/8&7/16&7/16&6/16&21/64&19/64
\\
\hline
${ST_n}$ & 3/4 & 3/4&5/8& 10/16&9/16&9/16&34/64&34/64
\\
\hline
${FC_n}$ & 3/4 &3/4&5/8&10/16&9/16&9/16&34/64&{34}/{64}
\\
\hline
\end{tabular}
\end{table}

Let us shortly discuss these results. 
First, note that 
{\it all} of the states in the table violate a Bell 
inequality of the type $|\mean{\BB}|\leq \CC(G)$ since
for all states and $n$ in Table I $\DD(G)<1$ holds. 
Then, it is remarkable, that long linear chains and 
large rings have a small $\DD$ (i.e. the violation
of the Bell inequality is large), while the violation
for the GHZ state is not so large. This in contrast 
to the usual Mermin inequality for several qubits. 
Due to Lemma 4, the values $\DD(ST_n)$ and 
$\DD(FC_n)$ always coincide. However, it is interesting that 
$\DD(FC_6)\not\leq\DD(FC_3)\DD(FC_3),$  
thus
a generalization of Lemma 3 to arbitrary connections
between the graphs is not true.  
In general, we have:

{\bf Theorem 1.} Any graph state violates local realism. 
\\
{\it Proof.} If the graph consists only of two connected 
vertices, the graph state is equivalent to a two qubit 
singlet state, which violates the original Bell 
inequality \cite{bell64}. Connected graphs with more
vertices always contain a subgraph with three vertices.
Due to Table 1 and the argumentation after Lemma 3
this implies that the graph state violates local realism.
$\qed$

{\bf Theorem 2.} Let ${G_1},{G_2},{G_3},...$
be a family of tree graphs 
(i.e., graphs which do not contain any closed ring)
with an increasing number of vertices such that 
each $G_i$ contains as a subgraph a linear
chain of a size which increases linearly with $i.$
Then violation of a Bell inequality for 
$\ket{G_i}$ increases exponentially with $i.$
\\
{\it Proof.} From Lemma 3 and Table I it follows
that for linear cluster graphs $LC_i$ the value $1/\DD(LC_i)$
increases exponentially. For tree 
graphs Lemma 3 can be applied again to show that
$\DD(G_i)$ is smaller than the $\DD$ from the longest
linear cluster in $G_i.$
$\qed$

Similar statements as the one expressed in Theorem 2 
can be derived for other families of graphs. In fact, 
all graphs which are built by connecting an increasing 
number of subgraphs via single edges exhibit an 
exponentially increasing violation of local realism, 
due to Lemma 3.

Let us compare the Bell inequalities with other entanglement 
properties. Here, our restriction to the 
observables $X,Y,Z$ becomes crucial.
A Bell inequality $|\mean{\BB}| \leq \CC(G)$ is
equivalent to a witness of the type
\be
\WW=\DD(G)-\ketbra{G},
\label{wit}
\ee
i.e., a quantum mechanical state violates the Bell 
inequality iff $\mean{\WW}<0$ \cite{remark2}. This has two 
consequences.
First, since all fully separable states 
$\ket{\psi}=\ket{a}\ket{b}...\ket{n}$
admit a LHV description, here 
$\DD(G)\geq 
\max_{\ket{a}\ket{b}...\ket{n}}|\braket{a,b,...,n}{G}|^2$
has to hold.  The quantity 
$1-\max_{\ket{a}\ket{b}...\ket{n}}|\braket{a,b,...,n}{G}|^2$
has been shown to be an entanglement monotone for multipartite
systems, the so-called {\it geometric measure of entanglement} 
\cite{wei}. So our bounds on $\CC$ also deliver lower bounds for 
this measure of entanglement for graph states. In turn, 
the fact that the geometric measure equals $1/2$ for all GHZ 
states implies that always $\DD(ST_n)\geq 1/2.$
Second, we can state: 

{\bf Theorem 3.}
If $\DD(G)\geq 1/2$ then the Bell inequality 
$|\mean{\BB}| \leq \CC(G)$ detects only states
which have a negative partial transpose with
respect to each partition.
\\
{\it Proof.} Let us fix a bipartite splitting
for the multipartite system. The graph state 
has a Schmidt decomposition 
$\ket{G}=\sum_{i=1}^k a_i \ket{ii}$ 
with respect to this splitting. It is known that then 
$1/k \leq a_0^2 \leq 1/2$ if $a_0$ is the biggest Schmidt 
coefficient \cite{GT04}. If we define
$\ket{\psi}=\sum_{i=1}^k \ket{ii}/\sqrt{k}$
it is also known that the witness
$\tilde{\WW}=\eins/k-\ketbra{\psi}$
detects only states which have a non-positive
partial transpose with 
respect to this partition \cite{schmidtwit}.
However the witness $\WW$ detects even less states, 
since
$
\WW-k {a_0}^2\tilde{\WW} = 
(\DD-a_0^2)\eins-\ketbra{G}+k a_0^2\ketbra{\psi}
\geq k a_0^2\ketbra{\psi}-\ketbra{G}\geq 0
$
which implies that $\mean{\WW}<0$ only if 
$\mean{\tilde{\WW}}<0.$ 
$\qed$

In conclusion, we have derived a family of Bell 
inequalities for multipartite systems based on the 
graph state formalism. These inequalities are 
maximally violated by graph states and allow to detect 
the non-locality of all  graph states. Also, 
the inequalities can be related to other topics as
the geometric measure of entanglement and the 
criterion of the partial transposition. The fact that
graph states do not admit a LHV model strongly 
suggests that tasks like measurement based quantum 
computation and quantum error correction cannot be 
described within classical physics.

We thank Marc Hein for discussions and P.H.~thanks the 
Universit\"{a}t Innsbruck for hospitality. 
This work has been supported by the DFG (Schwerpunkt 
Quanteninformationsverarbeitung)
and the EU (MEIF-CT-2003-500183, IST-2001-38877, -39227, 
OLAQI).

{\it Appendix ---}
Here we give the proof of Lemma 3. Let us consider 
a given LHV model and assume that the connection is 
between the vertices $i_0 \in G_1$ and $j_0 \in G_2.$ 
We can  write any stabilizing operator of $G$ as 
$s_{ij}(G)= a_i b_j$ where $a_i$ contains only 
$g_k$ with $k \in G_1$ and $b_j$ contains only 
$g_k$ belonging to $G_2.$ We can arrange the set 
$\{a_i\}=\fa_1\cup ... \cup \fa_6$ 
into six disjoint subsets in the following way: 
$\fa_1,\fa_2$ contain the $a_i$ 
where $g_{i_0}$ is absent, $\fa_3,\fa_4$ have 
an $X^{(i_0)}$ at the vertex $i_0$ and $\fa_5,\fa_6$ have 
a $Y^{(i_0)}.$ Note that this implies that
$\fa_3,\fa_4,\fa_5,$ and $\fa_6$ 
comprise the $a_i$ which contain $g_{i_0}.$ 
For $a_i \in \fa_1, \fa_3, \fa_5$ the LHV model gives
$\mean{a_i}=+1$ while for the $a_i \in \fa_2,\fa_4$ or  
$\fa_6$ we have $\mean{a_i}=-1.$ 
Let us denote the number of elements in the sets 
$\fa_1,\fa_2,\fa_3,\fa_4,\fa_5$ and $\fa_6$ by 
$p_a,q_a,r_a,s_a,t_a$ and $u_a.$ 
We can define $\gamma_a=:p_a-q_a, \delta_a:=r_a-s_a$
and $\varepsilon_a=t_a-u_a.$
Then we have 
$|\mean{\BB(G_1)}|=|\gamma_a+\delta_a+\varepsilon_a|\leq \CC(G_1).$
We can do the same ordering for the $b_j,$ introducing 
$\fb_1,...,\fb_6$ and $p_b,...,u_b$ and 
$\gamma_b,\delta_b,\varepsilon_b,$ with
$|\gamma_b+\delta_b+\varepsilon_b|\leq \CC(G_2).$

Let us investigate the relationships between  
$\gamma, \delta,$ and $ \varepsilon$ in some 
more detail. By flipping the sign which is assigned
to $X^{(i_0)}$ by the given LHV model we can construct
a new LHV model with 
$|\mean{\BB(G_1)}|=|\gamma_a-\delta_a+\varepsilon_a|
\leq \CC(G_1).$ We can also flip the signs of 
$Y^{(i_0)},X^{(j_0)}$ or $Y^{(j_0)}$ leading to 
new bounds of the type 
$|\gamma \pm \delta \pm \varepsilon| \leq \CC.$
All in all, this leads to the conclusion, that 
for all sixteen combinations of signs
\be
|(\gamma_a \pm \delta_a \pm \varepsilon_a) 
(\gamma_b \pm \delta_b \pm \varepsilon_b)| 
\leq \CC(G_1)\CC(G_2)
\label{toll3}
\ee
holds. Finally, note that
the operators $s_{ij}$ can be grouped into 36
groups according to $\fs_{ij}=\fa_i\fb_j.$
The mean values $\mean{s_{ij}}=\mean{a_i b_j}$ 
would factorize if there were no connection 
between the graphs. In this case, 
the Lemma is trivial.

What changes for the $\mean{s_{ij}}$ due to the 
extra connection? The $\mean{s_{ij}}$ can be written
in a $6 \times 6$ block matrix according to the 
grouping into the $\fs_{ij},$ where each block
bears the sign of the corresponding $\mean{s_{ij}}.$
In the blocks $\fs_{ij}$ with $i\leq 2$ or  
$j\leq 2$ the extra connection only 
introduces transformations of the type 
$Z\leftrightarrow\eins$ at $i_0$ or $j_0,$
which can be neglected due to Lemma 1.
More interesting is the $4 \times 4$ block matrix
of the blocks $\fs_{ij}$ with 
$3\leq i,j\leq 6.$ One can calculate that 
here the extra connection induces the 
transformation
$\{X^{(i_0)}X^{(j_0)}, X^{(i_0)}Y^{(j_0)}, Y^{(i_0)}X^{(j_0)}, 
Y^{(i_0)}Y^{(j_0)}\}
\mapsto
\{Y^{(i_0)}Y^{(j_0)}, -Y^{(i_0)}X^{(j_0)}, -X^{(i_0)}Y^{(j_0)}, 
X^{(i_0)}X^{(j_0)}\}$ 
on the vertices $i_0$ and $j_0.$ 
So, depending on the LHV model, this results in this block matrix 
in two possible changes of signs. They can be written 
in the following way:
\be
\left[
\begin{array}{cc|cc}
+ & - & + & - \\
-& + &- &+ \\
\hline
+ & - & + & - \\
-& + &- &+ \\
\end{array}
\right]
\mapsto
\left[
\begin{array}{cc|cc}
+ & - & - & + \\
-& + &+ &- \\
\hline 
- & + & + & - \\
+& - &- &+ \\
\end{array}
\right]
\mbox{ or }
\left[
\begin{array}{cc|cc}
- & + & + & - \\
+& - &- &+ \\
\hline 
+ & - & - & + \\
-& + &+ &- \\
\end{array}
\right].
\nonumber
\ee
For the  first possible transformation 
we have to show that
$
|\sum_{ij} \mean{s_{ij}} | = 
|\gamma_a (\gamma_b + \delta_b+  \varepsilon_b)
+\delta_a (\gamma_b + \delta_b - \varepsilon_b)
+\varepsilon_a (\gamma_b - \delta_b +\varepsilon_b)|
\leq \CC(G_1)\CC(G_2)
$
This can be derived from Eq.~(\ref{toll3}), distinguishing  
64 cases depending on the signs of 
$\gamma_a,\gamma_b,\delta_a,\delta_b,\varepsilon_a$ and $\varepsilon_b.$
For instance, if  $\gamma_a, \delta_a, \delta_b \geq 0$ 
and $\gamma_b,\varepsilon_a,\varepsilon_b<0$ we use
$
(\gamma_a + \delta_a + \varepsilon_a) 
(\gamma_b + \delta_b + \varepsilon_b) 
\leq
\gamma_a (\gamma_b + \delta_b+  \varepsilon_b)
+\delta_a (\gamma_b + \delta_b - \varepsilon_b)
+\varepsilon_a (\gamma_b - \delta_b +\varepsilon_b)
\equiv 
\sum_{ij} \mean{s_{ij}}
\leq
- (\gamma_a + \delta_a  -\varepsilon_a) 
(\gamma_b - \delta_b + \varepsilon_b)
,
$
yielding an upper and a lower bound for $\sum_{ij} \mean{s_{ij}}.$
The proof of the other 63 cases and the second transformation
is similar. 
$\qed$


\begin{thebibliography}{99}
\bibitem{remark2}
Note that in quantum theory always $\mean{\BB} \geq 0,$ 
since $\BB(G)=2^n \ketbra{G}$ is positive.
However, defining $\CC$ via the absolute value turns 
out to be convenient later.
\end{thebibliography}
\end{document}